\documentclass[11pt]{article} %
\usepackage{graphicx}
\usepackage{fancyhdr}
\usepackage{makeidx}
\usepackage{amssymb,amsmath}
\usepackage{physics}
\usepackage{gensymb}
\usepackage{upquote}
\usepackage{ulem}
\usepackage{xcolor}
\usepackage{cite}
\usepackage[utf8]{inputenc}
\usepackage[nottoc]{tocbibind}
\usepackage{setspace}
\usepackage{enumerate}
\usepackage{xr}
\usepackage[hidelinks]{hyperref}
\DeclareUnicodeCharacter{0308}{HERE!HERE!}
\DeclareUnicodeCharacter{030C}{HERE!HERE!}
\newcommand{\bd}{\begin{document}}
	\newcommand{\ed}{\end{document}}
\newcommand{\bc}{\begin{center}}
	\newcommand{\ec}{\end{center}}
\newcommand{\vs}{\vspace}
\newcommand{\hs}{\hspace}
\newcommand{\beq}{\begin{equation}}
\newcommand{\eeq}{\end{equation}}
\newcommand{\beqs}{\begin{eqn*}}
	\newcommand{\eeqs}{\end{eqn*}}
\newcommand{\bq}{\begin{quote}}
	\newcommand{\eq}{\end{quote}}
\newcommand{\lb}{\linebreak}
\newcommand{\mb}{\makebox}
\newcommand{\fb}{\framebox}
\newcommand{\mc}{\multicolumn}
\newcommand{\ben}{\begin{enumerate}}
	\newcommand{\een}{\end{enumerate}}
\newcommand{\bit}{\begin{itemize}}
	\newcommand{\eit}{\end{itemize}}
\newcommand{\ov}{\overline}
\newcommand{\un}{\underline}
\newcommand{\lt}{\left}
\newcommand{\rt}{\right}
\newcommand{\ba}{\begin{array}}
	\newcommand{\ea}{\end{array}}
\newcommand{\beqa}{\begin{eqnarray}}
\newcommand{\eeqa}{\end{eqnarray}}
\newcommand{\beqas}{\begin{eqnarray*}}
	\newcommand{\eeqas}{\end{eqnarray*}}
\newcommand{\bfg}{\begin{figure}}
	\newcommand{\efg}{\end{figure}}
\newcommand{\pad}{\partial}
\newcommand{\nn}{\nonumber}
\newcommand{\la}{\leftarrow}
\newcommand{\ra}{\rightarrow}
\newcommand{\lgla}{\longleftarrow}
\newcommand{\lgra}{\longrightarrow}
\newcommand{\La}{\Leftarrow}
\newcommand{\Ra}{\Rightarrow}
\newcommand{\Lra}{\Leftrightarrow}
\newcommand{\Lgla}{\Longleftarrow}
\newcommand{\Lgra}{\Longrightarrow}
\renewcommand{\a}{\alpha}
\renewcommand{\b}{\beta}
\newcommand{\g}{\gamma}
\newcommand{\G}{\Gamma}
\renewcommand{\d}{\delta}
\newcommand{\D}{\Delta}
\newcommand{\e}{\epsilon}
\newcommand{\eps}{\epsilon}
\newcommand{\s}{\sigma}
\newcommand{\m}{\mu}
\newcommand{\n}{\nu}
\renewcommand{\S}{\Sigma}
\newcommand{\p}{\pi}
\newcommand{\om}{\omega}
\newcommand{\Om}{\Omega}
\newcommand{\tri}{\triangle}
\newcommand{\ti}{\times}
\newcommand{\f}{\frac}
\newcommand{\ds}{\displaystyle}
\newcommand{\bm}[1]{\mb{{\boldmath $#1$}}}
\newcommand{\alter}[2]{\lt\{ \ba{ll}#1 \\ #2 \ea \rt.}
\newcommand{\alt}[4]{\lt\{ \ba{ll}#1 & \mb{if \, \,}#2 \\ #3 & \mb{}#4 \ea
	\rt.}
\newcommand{\altn}[4]{\lt\{ \ba{rl}#1 & \mb{if \, \,}#2 \\ #3 & \mb{}#4 \ea
	\rt.}
\newcommand{\altif}[4]{\lt\{ \ba{ll}#1 & \mb{if \, \,}#2 \\ #3 &
	\mb{if \, \,}#4 \ea \rt.}
\newcommand{\altnif}[4]{\lt\{ \ba{rl}#1 & \mb{if \, \,}#2 \\ #3 &
	\mb{if \, \,}#4 \ea \rt.}
\newcounter{algc}
\newcounter{romc}
\newcounter{Alphc}
\newcommand{\bl}{\begin{list}{{\it Step} ~\arabic{algc}~:} {\usecounter{algc}
			\setlength{\topsep}{0pt} \setlength{\itemsep}{0pt}}}
	\newcommand{\el}{\end{list}}
\newcommand{\blr}{\begin{list}{~\roman{romc}~:} {\usecounter{romc}
			\setlength{\topsep}{0pt} \setlength{\itemsep}{0pt}}}
	\newcommand{\elr}{\end{list}}
\newcommand{\bla}{\begin{list}{~\Alph{Alphc}~:} {\usecounter{Alphc}
			\setlength{\topsep}{0pt} \setlength{\itemsep}{0pt}}}
	\newcommand{\ela}{\end{list}}
\newcommand{\tsup}{\textsuperscript}
\newcommand{\tsub}{\textsubscript}
\newtheorem{theorem}{Theorem}
\makeatletter
\usepackage{lineno}
\usepackage{pdfpages}
\begin{document}
 \title{Simultaneously enhancing brightness and purity of WSe$_2$ single photon emitter using high-aspect-ratio nanopillar array on metal}
 \author{Mayank Chhaperwal$^{1}$, Himanshu Madhukar Tongale$^1$, Patrick Hays$^2$, Kenji Watanabe$^3$,\\ Takashi Taniguchi$^4$, Seth Ariel Tongay$^2$ and Kausik Majumdar$^{1*}$\\
	$^1$Department of Electrical Communication Engineering, \\Indian Institute of Science, Bangalore 560012, India\\
    $^2$Materials Science and Engineering, School for Engineering of Matter, Transport and Energy,\\ Arizona State University, Tempe, Arizona 85287, USA\\
	$^3$Research Center for Electronic and Optical Materials,\\ National Institute for Materials Science, 1-1 Namiki, Tsukuba 305-044, Japan\\
	$^4$Research Center for Materials Nanoarchitectonics,\\ National Institute for Materials Science, 1-1 Namiki, Tsukuba 305-044, Japan\\
	$^*$Corresponding author, email: kausikm@iisc.ac.in}
\date{}
\maketitle
\begin{abstract}
     Monolayer semiconductor transferred on nanopillar arrays provides site-controlled, on-chip single photon emission, which is a scalable light source platform for quantum technologies. However, the brightness of these emitters reported to date often falls short of the perceived requirement for such applications. Also, the single photon purity usually degrades as the brightness increases. Hence, there is a need for a design methodology to achieve enhanced emission rate while maintaining high single photon purity. Using WSe$_2$ on high-aspect-ratio ($\sim 3$ - at least two-fold higher than previous reports) nanopillar arrays, here we demonstrate $>10$ MHz single photon emission rate in the 770-800 nm band that is compatible with quantum memory and repeater networks (Rb-87-D1/D2 lines), and satellite quantum communication. The emitters exhibit excellent purity (even at high emission rates) and improved out-coupling due to the use of a gold back reflector that quenches the emission away from the nanopillar.
\end{abstract}
\textbf{keywords:} Single photon source, quantum emitter, WSe$_2$, 2D materials, brightness, second order correlation.\\

On-chip single photon emitters (SPEs) provide a scalable approach towards photonics-based quantum technologies such as quantum computing, quantum communication, and quantum metrology \cite{montblanch_layered_2023,turunen_quantum_2022}. SPEs based on transition metal dichalcogenides (TMDCs) are highly attractive as the two-dimensional nature of the host provides several possible advantages, such as, ease of integration with photonic/plasmonic cavities and waveguides\cite{luo_deterministic_2018,Cai2018,Peyskens2019,Iff2021,drawer_monolayer-based_2023,sortino2021}, low outcoupling loss \cite{drawer_monolayer-based_2023,luo_deterministic_2018,Iff2021,Peyskens2019}, gate induced spectral tunability \cite{Mukherjee2020,schwarz2016,Guo2023,stevens_enhancing_2022}, and the possibility of integration with electrical excitation scheme \cite{Guo2023,so2021,Palacios-Berraquero2016}. In nanopillar-based SPE implementations, the strain introduced on the monolayer through a nanopillar helps funnel excitons \cite{Johari2012,Shen2016,brooks_theory_2018,Harats2020,Niehues2018} toward the defect located at the pillar site, improving the net quantum efficiency of single photon emission \cite{branny2017,palacios-berraquero2017}. While such sources are highly promising due to the spatially deterministic nature of the array of quantum emitters at the lithographically defined positions \cite{sortino2021,parto2021,branny2017,Mukherjee2020,palacios-berraquero2017,Cai2018,luo2019},  there are two key challenges that must be addressed for practical quantum applications.

\begin{figure}[!hbt]
	\vs{-0.9in}
	\hs{-3.3in}
	\includegraphics[scale=1]{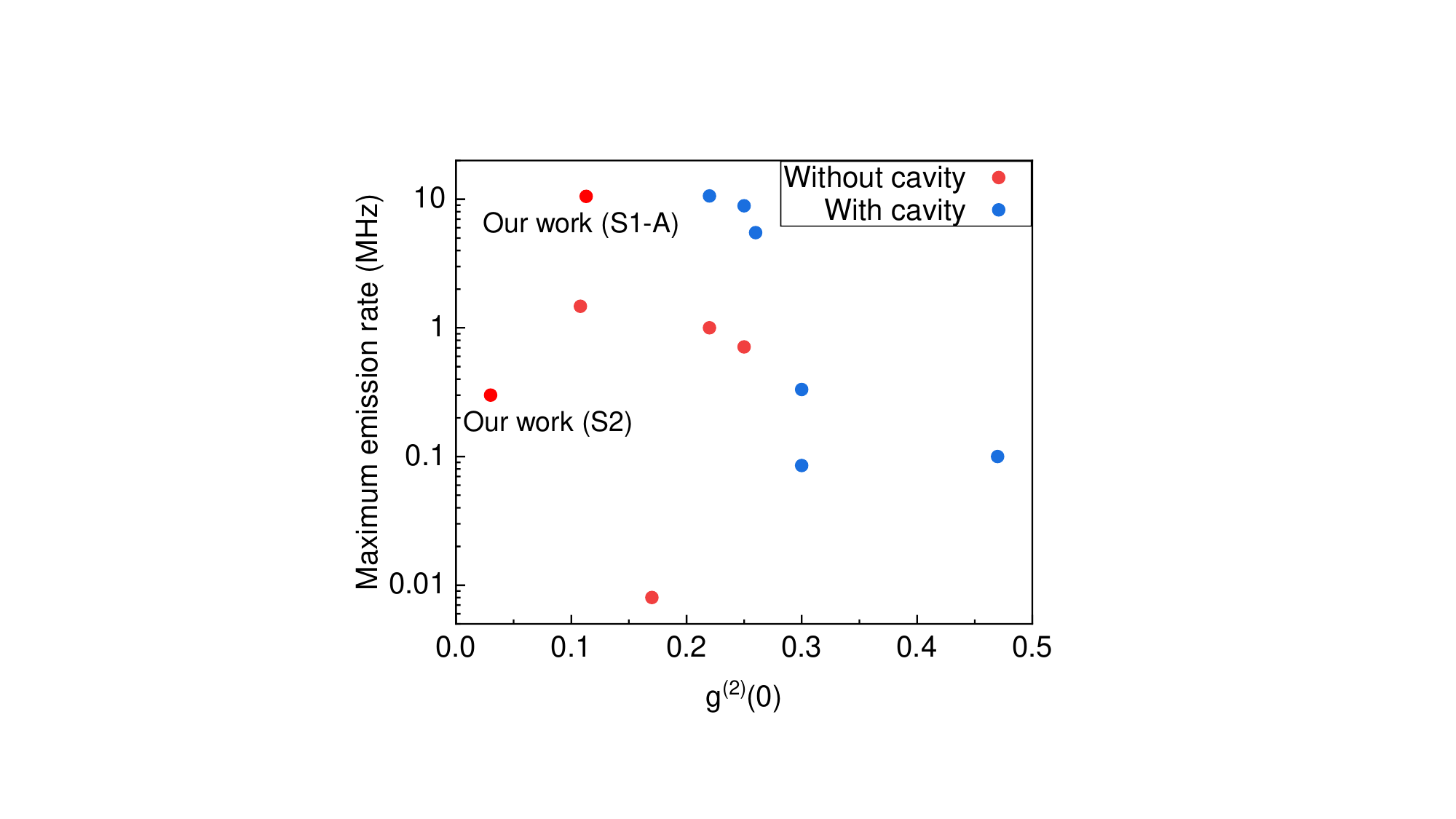}
	\vspace{-1in}
	\caption{\textbf{Status of TMDC-based single photon emitters and benchmarking.} Brightness (in terms of maximum collected emission rate in MHz) against single photon purity [in terms of $g^{(2)}(0)$] plotted for reported TMDC-based single photon emitters \cite{sortino2021,luo_deterministic_2018,so2021p,Kumar2015,Cai2018,Peyskens2019,flatten_microcavity_2018}. We have selected only those reports having scalable architecture and spatially deterministic SPE. Works implementing a photonic/plasmonic cavity are encoded in blue symbols, and those without cavities are encoded in red symbols. The plot also shows the results from this work having two different aspect ratios of pillars (samples S1-A and S2).}\label{fig:benchmarking}
\end{figure}

These two challenges are illustrated in \autoref{fig:benchmarking}, where we plot the measured brightness (in counts per second) as a function of single photon purity [in terms of $g^{(2)}(0)$] for reported TMDC-based SPEs \cite{sortino2021,luo_deterministic_2018,so2021p,Kumar2015,Cai2018,Peyskens2019,flatten_microcavity_2018}. The first observation is that the overall brightness is, in general, low, both with respect to other competing technologies (for example, quantum dots exhibiting several tens of MHz \cite{hoang_ultrafast_2016,sapienza_nanoscale_2015,aharonovich_solid-state_2016}) as well as the perceived requirement for several quantum technologies (in GHz, see \cite{gupta_single-photon_2023,aharonovich_solid-state_2016}). The emission intensity from TMDC-based SPEs usually saturates much earlier than the rate suggested by their lifetime. Auger annihilation has been suggested as the limiting factor for such early saturation  \cite{sortino2021}.  While methods, such as photonic/plasmonic cavities \cite{sortino2021,luo_deterministic_2018}, have improved the brightness of the SPEs, there is still a pressing need for further improvement.

While good single photon purity has been demonstrated in the past \cite{parto2021,branny2017,palacios-berraquero2017}, the second observation from \autoref{fig:benchmarking} is the lack of reports with simultaneous observation of low $g^{(2)}(0)$ and high emission rate. The diffraction-limited laser spot is much larger than the nanopillar diameter for optical excitation. This causes considerable contribution from the surrounding regions to the collected photon count. Unlike the SPE, the background emission is not bottlenecked by a single level and, hence, does not have a strong saturating behavior with increasing power. Accordingly, it often starts to dominate the SPE emission at higher excitation power, degrading the single photon purity.

The aim of this work is to bridge the above-mentioned gaps by introducing high-aspect-ratio nanopillars on metal film to significantly improve the brightness of the TMDC-based SPEs while maintaining an excellent single photon purity at high emission rates - a step towards practical SPE solution for various quantum technologies.\\

Transferring a monolayer on a nanopillar introduces a non-uniform strain in the flake, with  maximum strain at the pillar site. The band gap of the monolayer reduces with such local strain \cite{Johari2012,Shen2016,brooks_theory_2018}, causing excitons to funnel toward the pillar site. A point defect on the pillar captures these available excitons and emits subsequently, giving rise to single photon emission.

\begin{figure}[!hbt]
	\vs{-0.1in}
	\hs{-0.75in}
	\includegraphics[scale=0.57]{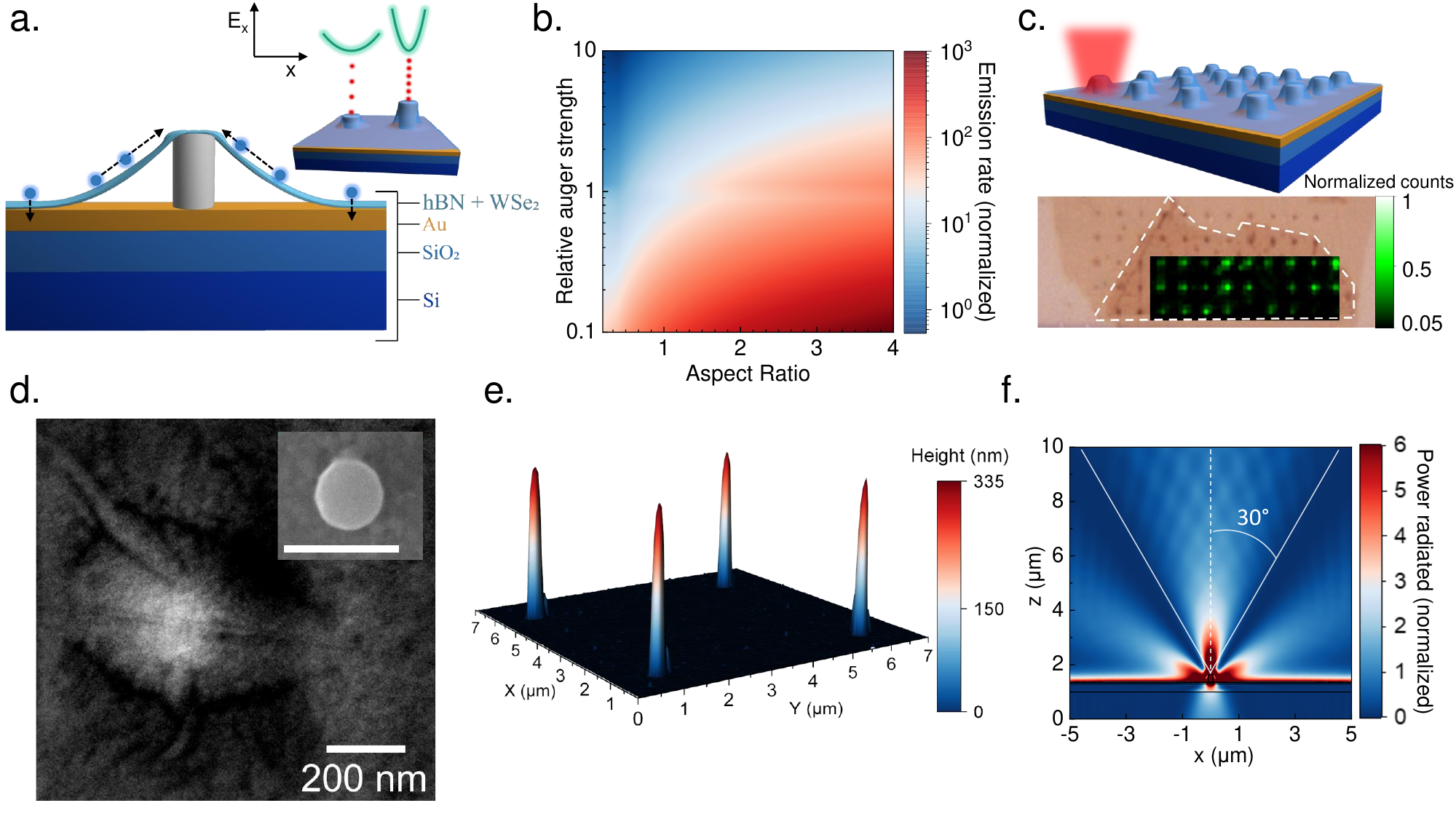}
	\vspace{-0.0in}
	\caption{\textbf{Design of the single photon emitter.} \textbf{a} Schematic cross section showing the different layers in the SPE. Inset: Schematic showing a comparison of shorter versus taller pillars with respect to their single photon emission rates. Excitonic bands showing different band-bending gradients are also shown. \textbf{b} Colour plot from our model showing the variation of normalized maximum emission rate achievable as a function of the nanopillar aspect ratio and Auger coefficient. \textbf{c} Top panel: Schematic showing the array of nanopillars covered with TMD monolayer and hBN. Individual nanopillars are spaced by $5$ $\mu$m such that the excitation laser spot covers only one of the nanopillars. Bottom panel: Optical image of the fabricated device with WSe\tsub2 monolayer (bounded by white dashed line) covering the array of nanopillars. Overlaid is the result from the PL map showing bright emission spots coinciding with the nanopillar locations. \textbf{d} SEM image of the WSe\tsub2 monolayer on a nanopillar. The nanopillar touches the gold-coated substrate around 250-300 nm away from the center of the nanopillar. The scale bar is 200 nm. Inset: SEM image of a bare nanopillar showing a diameter of around 110 nm (scale bar is 200 nm). \textbf{e} AFM image of a section of the fabricated nanopillar array showing a set of 4 nanopillars with height around 330 nm. \textbf{f} Result from the FDTD simulation showing the spatial distribution of the emission pattern of the SPE on the pillar. The solid white lines indicate the collection angle of our objective with an NA of 0.5.}\label{fig:Design and char}
\end{figure}

The overall design of our SPE array employing monolayer WSe\tsub2 on nanopillars has three important features,  as illustrated schematically in \autoref{fig:Design and char}a. The design is inspired by our model (discussed later) predicted brightness of the SPE as a function of Auger strength and pillar aspect ratio (height:diameter, denoted as $\alpha$), shown in \autoref{fig:Design and char}b. The results suggest that Auger-induced exciton annihilation in the strain-induced potential well on the nanopillar prevents the SPE from achieving its lifetime-limited emission rate. However, a higher $\alpha$ helps to overcome this limitation through stronger exciton funneling toward the pillar site (see inset of \autoref{fig:Design and char}a).  One drawback of high-aspect-ratio nanopillars made of hard material, such as SiO\tsub2, having a rough top surface, is the possibility of piercing through the flake during the transfer process \cite{palacios-berraquero2017,proscia_near-deterministic_2018}. To mitigate this, we fabricate the nanopillars using a negative photoresist followed by baking (see \textbf{Methods} in Supporting Information 1). The polymer-based negative resist provides a comparatively softer structure and a smoother top surface. We fabricate an array of such nanopillars, keeping a pillar-to-pillar spacing (5 $\mu$m) well beyond the excitation laser spot size ($\sim 1.5$ $\mu$m) so that only one pillar is excited at a time (top panel of \autoref{fig:Design and char}c for a schematic view and the bottom panel for an optical image). \autoref{fig:Design and char}d shows an SEM image of a single nanopillar with WSe\tsub2 monolayer on top. The inset is an SEM image of a bare nanopillar with a diameter of about 110 nm. AFM measurement indicates a height of about 330 nm (\autoref{fig:Design and char}e). This gives us  $\alpha\sim$ $3$, significantly higher than previous reports\cite{palacios-berraquero2017,branny2017,parto2021}.

Secondly, encapsulating TMDCs with hBN smoothens the inhomogeneous potential fluctuation and screens the inter-exciton interaction. This reduces the Auger coefficient significantly through suppressed interaction among locally trapped excitons \cite{Lee2022,hoshi2017,Chatterjee2022}. Accordingly, we transfer a few-layer hBN flake on top of the monolayer/nanopillar stack. However, no hBN layer is introduced between the monolayer and the nanopillar to avoid a possible reduction in the strain.

Finally, in order to improve single photon purity, we mitigate the issue of the laser excitation spot being much larger than the nanopillar area by fabricating the nanopillars on gold-coated substrate. The part of the monolayer away from the nanopillar touches the gold film (\autoref{fig:Design and char}a), and the emission from those regions is almost completely quenched due to non-radiative charge transfer to gold \cite{chaudhary2020}. This allows us to collect the emission only from the nanopillar region selectively. To support this claim, we perform a photoluminescence (PL) scan on the sample at 295 K around the free exciton emission energy (excited using 532 nm laser), and the integrated emission intensity is overlaid on the optical image (bottom panel of \autoref{fig:Design and char}c). The bright emission spots coincide with the location of the nanopillars in the optical image. A line cut from the PL map is presented in \hyperref[Suppl:Supp_metalPL]{Supporting Information 2}. The emission from the flake away from the nanopillar is almost completely quenched. On the other hand, excitons funnel to the nanopillar locations from the suspended regions, providing high brightness at the pillar site (schematic in \autoref{fig:Design and char}a). Most of the pillars covered by the monolayer show bright emission, suggesting high yield and minimal damage to the flake during the transfer process. Finally, the gold-coating acts as a back-reflector, improving the out-coupling of the emitted photons as depicted from the simulated emission pattern (\autoref{fig:Design and char}f) using 2D finite-difference time-domain (FDTD) method.

We refer to this SPE stack (with $\alpha=3$) as structure S1 and prepare three different samples (S1-A, S1-B, and S1-C) containing arrays of nanopillars.\\

\begin{figure}[!hbt]
	\vs{-0.1in}
	\hs{-0.75in}
	\includegraphics[scale=0.57]{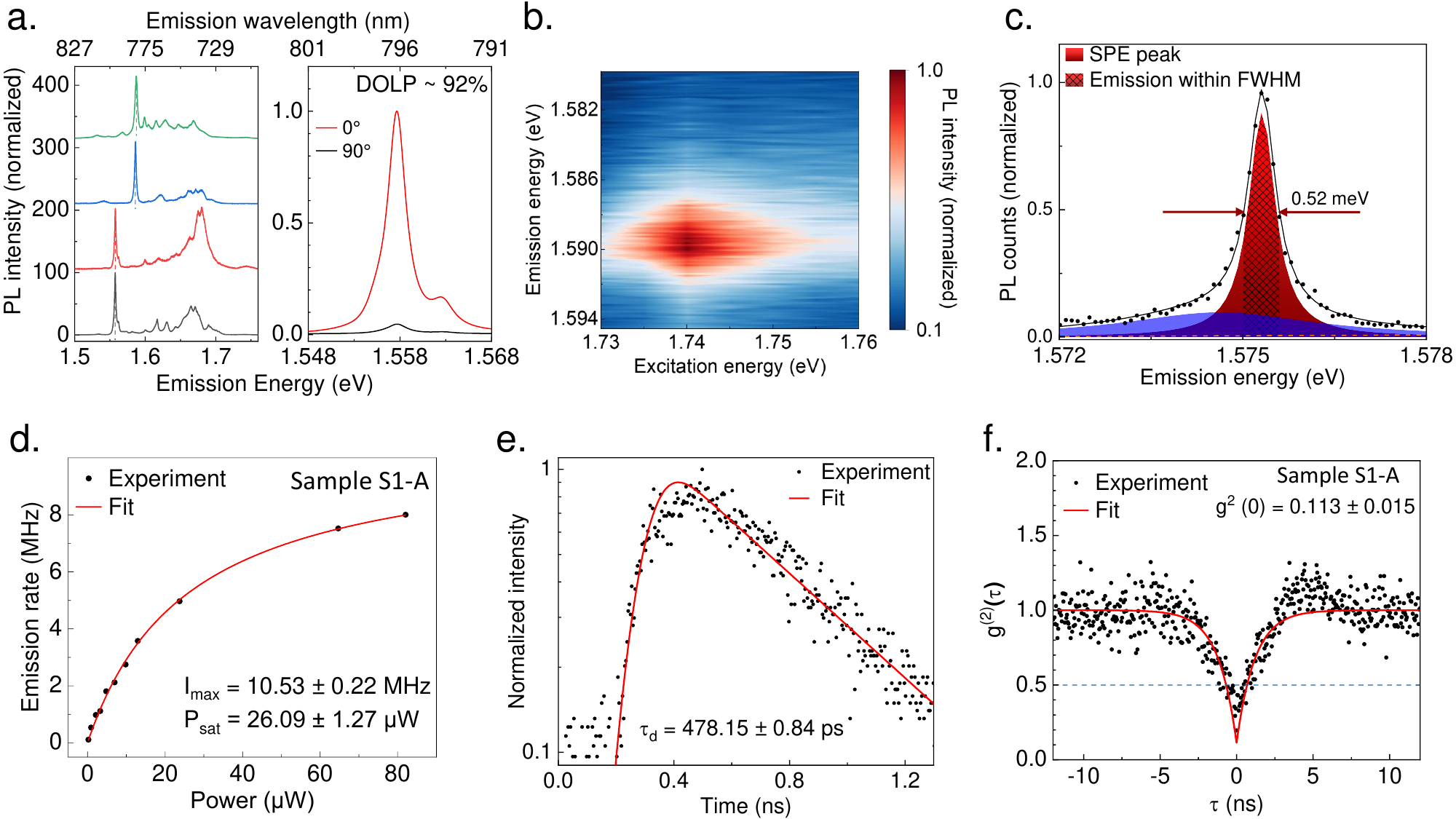}
	\vspace{-0.0in}
	\caption{\textbf{Characterization of the single photon emitter.} \textbf{a} Left panel: PL spectra from four different nanopillar sites. The spectral location of the SPEs is marked by dashed vertical lines. Right panel: Polarization resolved PL spectrum of the SPE peak showing a DOLP of $\sim$$92\%$. \textbf{b} Photoluminescence excitation measurement showing an enhancement in the SPE emission rate when the excitation is resonant to the free exciton energy of WSe$_2$. \textbf{c} Fitting of an SPE peak (data in black symbols and total fit in black solid trace) with Voigt functions and a background (orange dashed trace). The fitted peak (in red) is then integrated within its FWHM limits (black hatched portion) to calculate the integrated emission counts, thus removing the contribution from nearby peaks and background emission. The data shows a total measured line width of 520 $\mu$eV. \textbf{d} Integrated emission count rate from (\textbf{c}) plotted as a function of excitation power. The data fits with a saturation equation, giving a maximum emission rate of 10.53 $\pm$ 0.22 MHz and a saturation power of 26.09 $\mu$W. \textbf{e} TRPL results (data in black symbols and fitting in red trace) to characterize the formation time and the lifetime of the defect state. $t=0$ marks the location of the laser pulse firing. \textbf{f} Second-order correlation measurement to quantify the purity of the SPE (experimental data in symbols and fitting in the red trace).}\label{fig:optical_char}
\end{figure}

\autoref{fig:optical_char}a (left panel) shows representative PL emission spectra from four different nanopillars. The vertical dashed lines indicate the spectral position of the defect emission functioning as an SPE around 1.55 - 1.61 eV. The corresponding emission wavelength is in the 770-800 nm band and is thus suitable for quantum memory and repeater networks (being close to the Rb-87-D1/D2 line) \cite{simon_quantum_2010} and satellite communication applications\cite{gao_atomically-thin_2023}. Note that the SPE peak is red-shifted compared to the broad defect and multi-particle emission of WSe\tsub2 and, thus, is relatively free from background emission. \autoref{fig:optical_char}a (right panel) shows a high ($\sim$ 92\%) degree of linear polarization (DOLP) of these peaks. The DOLP value remains independent of the direction of the polarization of the excitation, suggesting the intrinsic linearly polarized nature of the SPE.

The SPE peak intensity strongly depends on the excitation wavelength. As an example, we choose an emitter around 1.59 eV and find that the peak intensity is highest when the excitation is resonant to the WSe$_2$ free exciton (\autoref{fig:optical_char}b).

The sharpest peak we measured has a full-width-at-half-maximum (FWHM) of 520 $\mu$eV and is shown in \autoref{fig:optical_char}c. We perform incident power ($P$) dependent PL measurement to characterize the brightness of the SPE. \autoref{fig:optical_char}c shows the method used to calculate the emission rate of the SPE. The peaks are fitted using a Voigt function, and the contributions from the side peaks and the background are removed. The SPE peak is then integrated within its FWHM (marked by the shaded portion) and then adjusted for our system efficiency of 5.1\%. This integrated emission intensity ($I$) of the SPE peak is plotted against $P$ in \autoref{fig:optical_char}d, and the data is fitted with  $I=I_{sat}\cdot\{P/(P+P_{sat})\}$. This gives us a maximum emission rate of 10.53 $\pm$ 0.22 MHz, which is free from other spurious contributions. Note that we deliberately do not report the collected SPAD count rate. This is because even after spectral filtering, the photons collected by the SPAD have a component arising from the background and side peaks. This component increases rapidly with the incident power and thus results in erroneous inflation of the measured emission rate of the device (\hyperref[Suppl:Supp_metalPL]{Supporting Information 3}). Power dependent emission rate from three other fabrication runs is shown in \hyperref[Suppl:Supp_MultiDevice]{Supporting Information 4}, suggesting a repeatable high emission rate from our samples.

Note that this is the emission rate collected by the objective [numerical aperture (NA) of 0.5] and is not corrected for the collection angle limited by its NA. Our FDTD simulation (\autoref{fig:Design and char}f) predicts that 33.93\% of the total photons emitted by the dipole above the substrate surface are collected by our objective. This gives an emission rate of $\sim$$31$ MHz above the surface of the substrate. Collection efficiency with respect to the total emission is 28.36\%, giving a total emission rate of $\sim$$37$ MHz. The FDTD simulation also rules out the existence of Purcell enhancement in the structure.

We perform time-resolved photoluminescence (TRPL) measurement to characterize the lifetime \cite{Palacios-Berraquero2016,peng2020,branny2017,wang2021,Peyskens2019,palacios-berraquero2017,shepard2017,Srivastava2015,Iff2021,Rosenberger2019,Kumar2015} of the trapped exciton in the quantum emitter (\autoref{fig:optical_char}e). After fitting with a decaying exponential convoluted with a Gaussian rise function (see \textbf{Methods} in Supporting Information 1), we extract a lifetime ($\tau_d$) of 487.15 $\pm$ 0.84 ps and a formation time ($\tau_f$) of 541 $\pm$ 0.95 ps.

To characterize the single photon nature of the emission, we perform second-order correlation measurement in a Hanbury Brown and Twiss (HBT) setup. After deconvolution from the Gaussian instrument response function (IRF) (see \textbf{Methods} in Supporting Information 1), we extract a $g^{(2)}(0)$ value of 0.113 $\pm$ 0.015 at a high collected count rate of 1 MHz. Note that the fit does not involve any background correction. The maintained single photon purity at this high emission rate confirms the functioning of the gold-coated substrate as an excellent background suppressor. Spectral separation of the SPE from typical broad defect emission of WSe\tsub2 also helps because the tail of such broad emission does not intrude into the spectral region of the SPE. The corresponding $g^{(2)}(\tau)$ plot for the sample exhibiting the narrowest SPE peak (spectrum in Figure \ref{fig:optical_char}c) is shown in \hyperref[Suppl:model_SPE]{Supporting Information 5}. Also, in \hyperref[Suppl:Supp_g2n]{Supporting Information 6}, we show $g^{(2)}(\tau)$ from a lower-aspect-ratio pillar (sample S2) with a $g^{(2)}(0)$ of 0.024 $\pm$ 0.017.

In \autoref{fig:benchmarking}, we compare our results with reported data \cite{sortino2021,luo_deterministic_2018,so2021p,Kumar2015,Cai2018,Peyskens2019,flatten_microcavity_2018} in the space of maximum emission rate versus $g^{(2)}(0)$, suggesting a superior emission rate while maintaining improved single photon purity. The emission rate from our high aspect-ratio pillars is 7-fold higher than the previous highest reported rate from pillars without cavities and is on par with works using photonic/plasmonic cavities. \\

\begin{figure}[!hbt]
	\vs{-0.1in}
	\hs{-0.75in}
	\includegraphics[scale=0.57]{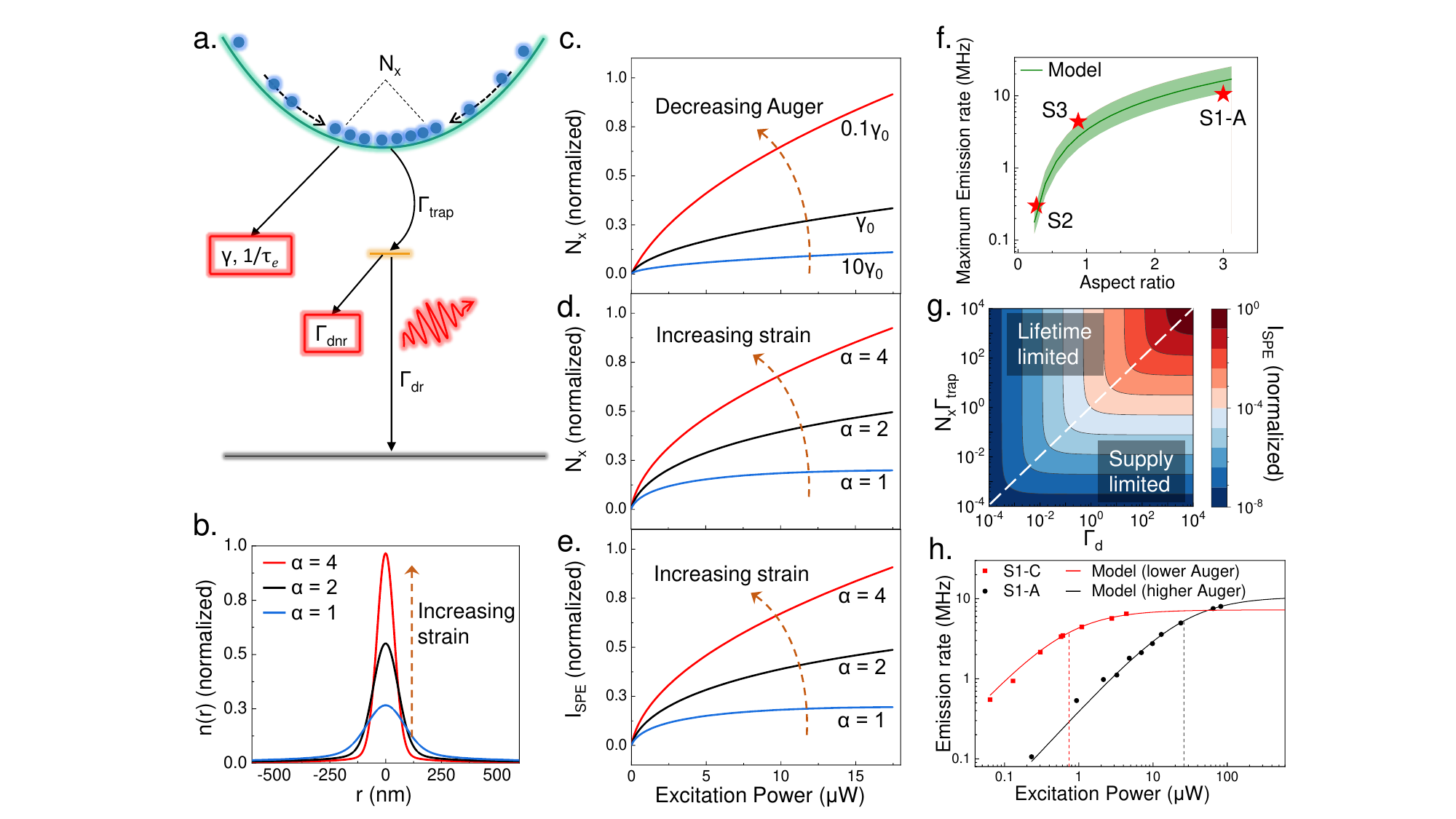}
	\vspace{-0.0in}
	\caption{\textbf{Single photon emitter model.} \textbf{a} Schematic diagram showing the mechanism of exciton funneling in the strain-induced potential well and capture by the defect state. The processes involved at different stages are also illustrated. \textbf{b} Exciton density as a function of radial distance from the center of the pillar with varying pillar aspect ratio ($\alpha$). \textbf{c-d} Exciton population near the center of the nanopillar ($N_x$) as a function of excitation power with (c) varying Auger coefficient relative to a base value $\gamma_0$ for a fixed aspect ratio of the pillar, and (d) varying aspect ratio of the pillar for a fixed Auger coefficient. \textbf{e} SPE emission intensity as a function of excitation power with a varying aspect ratio ($\alpha$) of the pillar. \textbf{f} Fitting of the model (in the green trace) with the experimentally measured maximum emission rate from samples S1-A, S2, and S3 (in red star symbols) as a function of pillar aspect ratio. The shaded portion demonstrates a $\pm$ 10 \% variation in the model parameters. \textbf{g} Maximum SPE emission intensity (normalized) achievable by varying the supply to the defect state ($N_x\Gamma_{trap}$) and total decay rate ($\Gamma_d$) of the defect state, for a given quantum efficiency. The dashed line indicates $\Gamma_d = N_x\Gamma_{trap}$ segregating the \textit{supply-limited} regime from the \textit{lifetime-limited} one. \textbf{h} Power dependent emission rate for S1-A (in black symbols) and S1-C (in red symbols) along with model fitting (solid traces). The fitted Auger coefficient for the red trace is about 100-fold smaller than the black trace. The dashed vertical lines indicate the corresponding saturation powers (0.74 $\mu$W for S1-C and 26.09 $\mu$W for S1-A). }\label{fig:model}
\end{figure}

We model the SPE as two interconnected subsystems (see \autoref{fig:model}a): (a) exciton funneling in the strain-induced potential well, and (b) these funneled excitons acting as a reservoir for exciting the defect-based atom-like two-level system, which then relaxes to the ground state, emitting single photons. We discuss these two parts of the model below.

The rate equation for the exciton density $n(r,t)$ at a radial distance $r$ from the center of the strain well is given by (see \hyperref[Suppl:ExcitonDynamics]{Supporting Information 7} for derivation):
\begin{equation}\label{eq:raten1_gamma}
\frac{\partial n(r,t)}{\partial t} = -\frac{\partial F(r)}{\partial r}+g(r)-\frac{n(r)}{\tau_e}-\gamma n^{2}(r)
\end{equation}
Here the first term on the right-hand side captures both the out-diffusion of excitons due to concentration gradient and the drift of excitons towards the pillar center due to strain gradient, $F$ being the total outward flux of the excitons. $g(r)$ is the generation rate with a spatial Gaussian profile, $\tau_e$ is the effective lifetime of the exciton (combining both radiative and non-radiative pathways, other than Auger process) and $\gamma$ is the Auger coefficient.

The calculated steady-state ($\frac{\partial n(r,t)}{\partial t}$ = 0) exciton density profile [$n(r)$] is plotted in \autoref{fig:model}b, suggesting stronger exciton funneling as $\alpha$ increases. In \autoref{fig:model}c, we plot the total exciton population [$N_x = \int_0^{r_0} 2\pi rn(r)dr$] near the center of the nanopillar as a function of $P$ for different $\gamma$ ($r_0$ is taken as 1 nm in the plots). The results indicate a saturation in $N_x$ with $P$, and $N_x$ saturates at a lower value for higher $\gamma$. This suggests that Auger annihilation, being a quadratic function, is the primary bottleneck to increase $N_x$. A higher $\alpha$ that induces stronger exciton funneling could be a way to mitigate such Auger-mediated saturation, as illustrated in  \autoref{fig:model}d.

An exciton from the reservoir that is in the vicinity is captured by a point defect, populating an atom-like two-level system consisting of the defect state and the ground state (\autoref{fig:model}a). The SPE emission energy is $\sim$ 200 meV lower than the free exciton (\autoref{fig:optical_char}a). The hybridization of this defect state with the dark exciton band \cite{hernandez_lopez_strain_2022,abramov_photoluminescence_2023,moon_strain-correlated_2020,xu_conversion_2023,linhart_localized_2019} is thus unlikely in our samples because of the large energy separation between them.

The steady-state emission rate of the SPE is given by (see \hyperref[Suppl:model_SPE]{Supporting Information 9} for derivation):
\begin{equation}\label{eq:I_defect}
I_{SPE} = \Gamma_{dr} \bigg(\frac{N_x\Gamma_{trap}}{N_x\Gamma_{trap} + \Gamma_{d}}\bigg)
\end{equation}
Here $\Gamma_{trap}$ quantifies the rate of capture of an individual exciton by the defect. $\Gamma_{d}=\frac{1}{\tau_d}=\Gamma_{dr}+\Gamma_{dnr}$ is the effective rate of relaxation from the defect state, considering both radiative ($\Gamma_{dr}$) and non-radiative ($\Gamma_{dnr}$) pathways.
In \autoref{fig:model}e, the calculated $I_{SPE}$ is plotted against incident power for varying $\alpha$, indicating an effective enhancement of the maximum emission rate with pillar aspect ratio. The variation of the calculated maximum emission rate is shown earlier (\autoref{fig:Design and char}b) in a colour plot as a function of the Auger coefficient and pillar aspect ratio.

To verify that such a high emission rate results from the increased strain due to the higher aspect ratio of our nanopillars, we prepare two other samples (S2 and S3) with a lower aspect ratio of $\alpha=0.27$ and $\alpha=0.88$ respectively. The maximum collected rate measured from these devices is plotted in \autoref{fig:model}f. The model provides an excellent fit to the experimentally observed maximum emission rate as a function of $\alpha$, verifying the validity of the model.

Note that $N_x\Gamma_{trap}$ provides a good estimate of the rate of formation of the defect-bound exciton (see \hyperref[Suppl:model_SPE]{Supporting Information 8-9}), and hence $N_x\Gamma_{trap}\approx\frac{1}{\tau_f}$. With this approximation, \autoref{eq:I_defect} reduces to the expected form:
\begin{equation}\label{eq:I_restate}
I_{SPE}=\eta \times\frac{1}{\tau_f+\tau_d}
\end{equation}
Here $\eta=\frac{\Gamma_{dr}}{\Gamma_{d}}$ is the quantum efficiency of the defect emission.

Equations \ref{eq:I_defect} and \ref{eq:I_restate} provide key insights into the possible ways to enhance the SPE emission rate, and the results are shown in \autoref{fig:model}g. For example, when $N_x\Gamma_{trap}\ll\Gamma_{d}$, that is, when the defect state is starved of exciton supply from the reservoir (lower half in \autoref{fig:model}g, denoted as \textit{`supply-limited'}),  $I_{SPE}\approx \eta N_x\Gamma_{trap} \approx \eta/\tau_f$ and hence $I_{SPE}$ exactly follows $N_x$. Since $N_x$ saturates at higher excitation power, $I_{SPE}$ also saturates prematurely without reaching the true lifetime limited rate of $\Gamma_{dr}$. Such saturation behavior thus does not arise from the limit of a 2-level quantum system. We believe a majority of the reported TMDC-based SPEs to date fall under this regime. Mitigating the Auger-induced early saturation by increasing the nanopillar aspect ratio can improve the emission rate in this regime (see \autoref{fig:model}e).

On the other hand, when $N_x\Gamma_{trap}\gg\Gamma_{d}$, that is, when the defect state is not starved by the supply of excitons (upper half in \autoref{fig:model}g, denoted as \textit{`lifetime-limited'}), $I_{SPE}\approx \Gamma_{dr}$ indicating radiative lifetime-limited emission rate, and thus further increasing the supply through enhanced pillar aspect ratio or reducing Auger recombination is not effective anymore in this regime. This also means that with adequate supply, $I_{SPE}$ can achieve $\Gamma_{dr}$ irrespective of the value of $\Gamma_{dnr}$.

To verify the model, in \hyperref[Suppl:Supp_LowerAuger]{Supporting Information 10}, we compare the emission spectra of samples S1-A and S1-C. S1-C uses the same aspect ratio pillars ($\alpha=3$) as S1-A, however, a different batch of WSe$_2$ that exhibits relatively weak background emission. This suggests a smoother potential profile and, thus, likely suppressed Auger loss due to reduced exciton localization \cite{Lee2022,hoshi2017,Chatterjee2022}. The defect peak emission is slightly blue shifted to 1.6 eV in this sample. Interestingly, we achieve a similar maximum count rate of 7.4 $\pm$ 0.17 MHz; however, the saturation excitation power ($P_{sat}$) is $\sim 35$-fold lower compared with S1-A. By varying the Auger coefficient for S1-A and S1-C, our model provides an excellent fit (\autoref{fig:model}h) with the experimentally obtained emission rate as a function of excitation power for both the samples.

The second factor on the right side in \autoref{eq:I_defect}, that is, $\beta \equiv \frac{N_x\Gamma_{trap}}{N_x\Gamma_{trap} + \Gamma_{d}}$ provides a quantitative estimate of the degree of supply limitation, with $\beta$ close to zero indicating strong supply limitation, and $\beta$ close to unity indicating lifetime-limited regime. In our sample S1-A with a high pillar aspect ratio, the extracted values of $\tau_f$ and $\tau_d$ from TRPL suggest $\beta\approx 0.47$. This justifies the lack of a proportionate increment in the maximum emission rate in S1-C (with respect to S1-A) in spite of significant suppression of the Auger coefficient.

Finally, with the estimated total emission rate of $\sim$$37$ MHz and $\beta\approx 0.47$, \autoref{eq:I_defect} suggests a radiative rate ($\Gamma_{dr}$) of $\sim$$79$ MHz and hence a radiative lifetime of $\sim$$12.6$ ns. Thus, there is a scope for further optimization of the pillar aspect ratio to improve the maximum emission rate by nearly a factor of two. However, a significant enhancement in the emission rate beyond this would require improving $\Gamma_{dr}$, possibly through cavity and material design, which could lead to scalable and highly pure single photon emission with rates beyond 100 MHz. Further, in the current work, there is a lack of precise control on the SPE emission wavelength, which could be addressed in the future through an external electric field-induced Stark shift.

\section*{Supporting Information}
The Supporting Information is available free of charge at XXX on fabrication and characterization methods, line cut from PL mapping of emitters,  SPAD count rate versus pure SPE count rate, emission rate data from additional samples, single photon purity data from additional samples, description of the model, a description of two-level emitter, SPE model details, comparison of devices with higher and lower Auger coefficients.

\section*{Acknowledgements}
K.M. acknowledges useful discussion with Naresh Babu Pendyala from ISRO. This work was supported in part by a Core Research Grant from the Science and Engineering Research Board (SERB) under Department of Science and Technology (DST), grants from Indian Space Research Organization (ISRO), a grant under SERB TETRA, a grant from I-HUB QTF, IISER Pune, and a seed funding under Quantum Research Park (QuRP) from Karnataka Innovation and Technology Society (KITS), K-Tech, Government of Karnataka. K.W. and T.T. acknowledge support from the JSPS KAKENHI (Grant Numbers 21H05233 and 23H02052) and World Premier International Research Center Initiative (WPI), MEXT, Japan. S. T. acknowledges direct support from DOE-SC0020653 (materials synthesis), NSF ECCS 2052527, DMR 2111812, and CMMI 2129412. The use of facilities within the Eyring Materials Center at Arizona State University is partly supported by NNCI-ECCS-1542160.

\section*{Competing Interests}
The authors declare no competing financial or non-financial interests.
\section*{Data Availability}
Data available on reasonable request from the corresponding author.\\
\bibliographystyle{unsrt}
\bibliography{citations}

\begin{thebibliography}{10}

\bibitem{montblanch_layered_2023}
Alejandro R.-P. Montblanch, Matteo Barbone, Igor Aharonovich, Mete Atatüre, and Andrea~C. Ferrari.
\newblock Layered materials as a platform for quantum technologies.
\newblock {\em \href{https://www.nature.com/articles/s41565-023-01354-x}{Nature Nanotechnology}}, 18(6):555, June 2023.

\bibitem{turunen_quantum_2022}
Mikko Turunen, Mauro Brotons-Gisbert, Yunyun Dai, Yadong Wang, Eleanor Scerri, Cristian Bonato, Klaus~D. Jöns, Zhipei Sun, and Brian~D. Gerardot.
\newblock Quantum photonics with layered {2D} materials.
\newblock {\em \href{https://www.nature.com/articles/s42254-021-00408-0}{Nature Reviews Physics}}, 4(4):219, January 2022.

\bibitem{luo_deterministic_2018}
Yue Luo, Gabriella~D. Shepard, Jenny~V. Ardelean, Daniel~A. Rhodes, Bumho Kim, Katayun Barmak, James~C. Hone, and Stefan Strauf.
\newblock Deterministic coupling of site-controlled quantum emitters in monolayer {WSe$_{\textrm{2}}$} to plasmonic nanocavities.
\newblock {\em \href{https://www.nature.com/articles/s41565-018-0275-z}{Nature Nanotechnology}}, 13(12):1137, December 2018.

\bibitem{Cai2018}
Tao Cai, Je-Hyung Kim, Zhili Yang, Subhojit Dutta, Shahriar Aghaeimeibodi, and Edo Waks.
\newblock Radiative {Enhancement} of {Single} {Quantum} {Emitters} in {WSe}$_{\textrm{2}}$ {Monolayers} {Using} {Site}-{Controlled} {Metallic} {Nanopillars}.
\newblock {\em \href{https://pubs.acs.org/doi/10.1021/acsphotonics.8b00580}{ACS Photonics}}, 5(9):3466, September 2018.

\bibitem{Peyskens2019}
Frédéric Peyskens, Chitraleema Chakraborty, Muhammad Muneeb, Dries Van~Thourhout, and Dirk Englund.
\newblock Integration of single photon emitters in {2D} layered materials with a silicon nitride photonic chip.
\newblock {\em \href{https://www.nature.com/articles/s41467-019-12421-0}{Nature Communications}}, 10(1):4435, September 2019.

\bibitem{Iff2021}
Oliver Iff, Quirin Buchinger, Magdalena Moczała-Dusanowska, Martin Kamp, Simon Betzold, Marcelo Davanco, Kartik Srinivasan, Sefaattin Tongay, Carlos Antón-Solanas, Sven Höfling, and Christian Schneider.
\newblock Purcell-{Enhanced} {Single} {Photon} {Source} {Based} on a {Deterministically} {Placed} {WSe}$_{\textrm{2}}$ {Monolayer} {Quantum} {Dot} in a {Circular} {Bragg} {Grating} {Cavity}.
\newblock {\em \href{https://pubs.acs.org/doi/10.1021/acs.nanolett.1c00978}{Nano Letters}}, 21(11):4715, June 2021.

\bibitem{drawer_monolayer-based_2023}
Jens-Christian Drawer, Victor~Nikolaevich Mitryakhin, Hangyong Shan, Sven Stephan, Moritz Gittinger, Lukas Lackner, Bo~Han, Gilbert Leibeling, Falk Eilenberger, Rounak Banerjee, Sefaattin Tongay, Kenji Watanabe, Takashi Taniguchi, Christoph Lienau, Martin Silies, Carlos Anton-Solanas, Martin Esmann, and Christian Schneider.
\newblock Monolayer-{Based} {Single}-{Photon} {Source} in a {Liquid}-{Helium}-{Free} {Open} {Cavity} {Featuring} 65\% {Brightness} and {Quantum} {Coherence}.
\newblock {\em \href{https://pubs.acs.org/doi/10.1021/acs.nanolett.3c02584}{Nano Letters}}, 23(18):8683, September 2023.

\bibitem{sortino2021}
Luca Sortino, Panaiot~G. Zotev, Catherine~L. Phillips, Alistair~J. Brash, Javier Cambiasso, Elena Marensi, A.~Mark Fox, Stefan~A. Maier, Riccardo Sapienza, and Alexander~I. Tartakovskii.
\newblock Bright single photon emitters with enhanced quantum efficiency in a two-dimensional semiconductor coupled with dielectric nano-antennas.
\newblock {\em \href{https://www.nature.com/articles/s41467-021-26262-3}{Nature Communications}}, 12(1):6063, October 2021.

\bibitem{Mukherjee2020}
Arunabh Mukherjee, Chitraleema Chakraborty, Liangyu Qiu, and A.~Nick Vamivakas.
\newblock {Electric field tuning of strain-induced quantum emitters in WSe$_{\textrm{2}}$}.
\newblock {\em \href{https://doi.org/10.1063/5.0010395}{AIP Advances}}, 10(7):075310, 07 2020.

\bibitem{schwarz2016}
S~Schwarz, A~Kozikov, F~Withers, J~K Maguire, A~P Foster, S~Dufferwiel, L~Hague, M~N Makhonin, L~R Wilson, A~K Geim, K~S Novoselov, and A~I Tartakovskii.
\newblock Electrically pumped single-defect light emitters in {WSe}$_{\textrm{2}}$.
\newblock {\em \href{https://iopscience.iop.org/article/10.1088/2053-1583/3/2/025038}{2D Materials}}, 3(2):025038, June 2016.

\bibitem{Guo2023}
Shi Guo, Savvas Germanis, Takashi Taniguchi, Kenji Watanabe, Freddie Withers, and Isaac~J. Luxmoore.
\newblock Electrically {Driven} {Site}-{Controlled} {Single} {Photon} {Source}.
\newblock {\em \href{https://pubs.acs.org/doi/10.1021/acsphotonics.3c00097}{ACS Photonics}}, 10(8):2549, August 2023.

\bibitem{stevens_enhancing_2022}
Christopher~E. Stevens, Hsun-Jen Chuang, Matthew~R. Rosenberger, Kathleen~M. McCreary, Chandriker~Kavir Dass, Berend~T. Jonker, and Joshua~R. Hendrickson.
\newblock Enhancing the {Purity} of {Deterministically} {Placed} {Quantum} {Emitters} in {Monolayer} {WSe}$_{\textrm{2}}$.
\newblock {\em \href{https://pubs.acs.org/doi/10.1021/acsnano.2c08553}{ACS Nano}}, 16(12):20956, December 2022.

\bibitem{so2021}
Jae-Pil So, Ha-Reem Kim, Hyeonjun Baek, Kwang-Yong Jeong, Hoo-Cheol Lee, Woong Huh, Yoon~Seok Kim, Kenji Watanabe, Takashi Taniguchi, Jungkil Kim, Chul-Ho Lee, and Hong-Gyu Park.
\newblock Electrically driven strain-induced deterministic single-photon emitters in a van der {Waals} heterostructure.
\newblock {\em \href{https://www.science.org/doi/10.1126/sciadv.abj3176}{Science Advances}}, 7(43):eabj3176, October 2021.

\bibitem{Palacios-Berraquero2016}
Carmen Palacios-Berraquero, Matteo Barbone, Dhiren~M. Kara, Xiaolong Chen, Ilya Goykhman, Duhee Yoon, Anna~K. Ott, Jan Beitner, Kenji Watanabe, Takashi Taniguchi, Andrea~C. Ferrari, and Mete Atatüre.
\newblock Atomically thin quantum light-emitting diodes.
\newblock {\em \href{https://www.nature.com/articles/ncomms12978}{Nature Communications}}, 7(1):12978, September 2016.

\bibitem{Johari2012}
Priya Johari and Vivek~B. Shenoy.
\newblock Tuning the {Electronic} {Properties} of {Semiconducting} {Transition} {Metal} {Dichalcogenides} by {Applying} {Mechanical} {Strains}.
\newblock {\em \href{https://pubs.acs.org/doi/10.1021/nn301320r}{ACS Nano}}, 6(6):5449, June 2012.

\bibitem{Shen2016}
Tingting Shen, Ashish~V. Penumatcha, and Joerg Appenzeller.
\newblock Strain {Engineering} for {Transition} {Metal} {Dichalcogenides} {Based} {Field} {Effect} {Transistors}.
\newblock {\em \href{https://pubs.acs.org/doi/10.1021/acsnano.6b01149}{ACS Nano}}, 10(4):4712, April 2016.

\bibitem{brooks_theory_2018}
Matthew Brooks and Guido Burkard.
\newblock Theory of strain-induced confinement in transition metal dichalcogenide monolayers.
\newblock {\em \href{https://link.aps.org/doi/10.1103/PhysRevB.97.195454}{Physical Review B}}, 97(19):195454, May 2018.

\bibitem{Harats2020}
Moshe~G. Harats, Jan~N. Kirchhof, Mengxiong Qiao, Kyrylo Greben, and Kirill~I. Bolotin.
\newblock Dynamics and efficient conversion of excitons to trions in non-uniformly strained monolayer {WS$_{\textrm{2}}$}.
\newblock {\em \href{https://www.nature.com/articles/s41566-019-0581-5}{Nature Photonics}}, 14(5):324, May 2020.

\bibitem{Niehues2018}
Iris Niehues, Robert Schmidt, Matthias Drüppel, Philipp Marauhn, Dominik Christiansen, Malte Selig, Gunnar Berghäuser, Daniel Wigger, Robert Schneider, Lisa Braasch, Rouven Koch, Andres Castellanos-Gomez, Tilmann Kuhn, Andreas Knorr, Ermin Malic, Michael Rohlfing, Steffen Michaelis De~Vasconcellos, and Rudolf Bratschitsch.
\newblock Strain {Control} of {Exciton}–{Phonon} {Coupling} in {Atomically} {Thin} {Semiconductors}.
\newblock {\em \href{https://pubs.acs.org/doi/10.1021/acs.nanolett.7b04868}{Nano Letters}}, 18(3):1751, March 2018.

\bibitem{branny2017}
Artur Branny, Santosh Kumar, Raphaël Proux, and Brian~D Gerardot.
\newblock Deterministic strain-induced arrays of quantum emitters in a two-dimensional semiconductor.
\newblock {\em \href{https://www.nature.com/articles/ncomms15053}{Nature Communications}}, 8(1):15053, May 2017.

\bibitem{palacios-berraquero2017}
Carmen Palacios-Berraquero, Dhiren~M. Kara, Alejandro R.-P. Montblanch, Matteo Barbone, Pawel Latawiec, Duhee Yoon, Anna~K. Ott, Marko Loncar, Andrea~C. Ferrari, and Mete Atatüre.
\newblock Large-scale quantum-emitter arrays in atomically thin semiconductors.
\newblock {\em \href{https://www.nature.com/articles/ncomms15093}{Nature Communications}}, 8(1):15093, May 2017.

\bibitem{parto2021}
Kamyar Parto, Shaimaa~I. Azzam, Kaustav Banerjee, and Galan Moody.
\newblock Defect and strain engineering of monolayer {WSe$_{\textrm{2}}$} enables site-controlled single-photon emission up to 150 {K}.
\newblock {\em \href{https://www.nature.com/articles/s41467-021-23709-5}{Nature Communications}}, 12(1):3585, June 2021.

\bibitem{luo2019}
Yue Luo, Na~Liu, Xiangzhi Li, James~C Hone, and Stefan Strauf.
\newblock Single photon emission in {WSe}$_{\textrm{2}}$ up 160 {K} by quantum yield control.
\newblock {\em \href{https://iopscience.iop.org/article/10.1088/2053-1583/ab15fe}{2D Materials}}, 6(3):035017, May 2019.

\bibitem{so2021p}
Jae-Pil So, Kwang-Yong Jeong, Jung~Min Lee, Kyoung-Ho Kim, Soon-Jae Lee, Woong Huh, Ha-Reem Kim, Jae-Hyuck Choi, Jin~Myung Kim, Yoon~Seok Kim, Chul-Ho Lee, SungWoo Nam, and Hong-Gyu Park.
\newblock Polarization {Control} of {Deterministic} {Single}-{Photon} {Emitters} in {Monolayer} {WSe}$_{\textrm{2}}$.
\newblock {\em \href{https://pubs.acs.org/doi/10.1021/acs.nanolett.1c00078}{Nano Letters}}, 21(3):1546, February 2021.

\bibitem{Kumar2015}
S.~Kumar, A.~Kaczmarczyk, and B.~D. Gerardot.
\newblock Strain-{Induced} {Spatial} and {Spectral} {Isolation} of {Quantum} {Emitters} in {Mono}- and {Bilayer} {WSe}$_{\textrm{2}}$.
\newblock {\em \href{https://pubs.acs.org/doi/10.1021/acs.nanolett.5b03312}{Nano Letters}}, 15(11):7567, November 2015.

\bibitem{flatten_microcavity_2018}
L.~C. Flatten, L.~Weng, A.~Branny, S.~Johnson, P.~R. Dolan, A.~A.~P. Trichet, B.~D. Gerardot, and J.~M. Smith.
\newblock Microcavity enhanced single photon emission from two-dimensional {WSe$_{\textrm{2}}$}.
\newblock {\em \href{https://pubs.aip.org/apl/article/112/19/191105/35374/Microcavity-enhanced-single-photon-emission-from}{Applied Physics Letters}}, 112(19):191105, May 2018.

\bibitem{hoang_ultrafast_2016}
Thang~B. Hoang, Gleb~M. Akselrod, and Maiken~H. Mikkelsen.
\newblock Ultrafast {Room}-{Temperature} {Single} {Photon} {Emission} from {Quantum} {Dots} {Coupled} to {Plasmonic} {Nanocavities}.
\newblock {\em \href{https://pubs.acs.org/doi/10.1021/acs.nanolett.5b03724}{Nano Letters}}, 16(1):270, January 2016.

\bibitem{sapienza_nanoscale_2015}
Luca Sapienza, Marcelo Davanço, Antonio Badolato, and Kartik Srinivasan.
\newblock Nanoscale optical positioning of single quantum dots for bright and pure single-photon emission.
\newblock {\em \href{https://www.nature.com/articles/ncomms8833}{Nature Communications}}, 6(1):7833, July 2015.

\bibitem{aharonovich_solid-state_2016}
Igor Aharonovich, Dirk Englund, and Milos Toth.
\newblock Solid-state single-photon emitters.
\newblock {\em \href{https://www.nature.com/articles/nphoton.2016.186}{Nature Photonics}}, 10(10):631, October 2016.

\bibitem{gupta_single-photon_2023}
Sunny Gupta, Wenjing Wu, Shengxi Huang, and Boris~I. Yakobson.
\newblock Single-{Photon} {Emission} from {Two}-{Dimensional} {Materials}, to a {Brighter} {Future}.
\newblock {\em \href{https://pubs.acs.org/doi/10.1021/acs.jpclett.2c03674}{The Journal of Physical Chemistry Letters}}, 14(13):3274, April 2023.

\bibitem{proscia_near-deterministic_2018}
Nicholas~V. Proscia, Zav Shotan, Harishankar Jayakumar, Prithvi Reddy, Charles Cohen, Michael Dollar, Audrius Alkauskas, Marcus Doherty, Carlos~A. Meriles, and Vinod~M. Menon.
\newblock Near-deterministic activation of room-temperature quantum emitters in hexagonal boron nitride.
\newblock {\em \href{https://opg.optica.org/abstract.cfm?URI=optica-5-9-1128}{Optica}}, 5(9):1128, September 2018.

\bibitem{Lee2022}
Yongjun Lee, Trang~Thu Tran, Youngbum Kim, Shrawan Roy, Takashi Taniguchi, Kenji Watanabe, Joon~I. Jang, and Jeongyong Kim.
\newblock Enhanced {Radiative} {Exciton} {Recombination} in {Monolayer} {WS}$_{\textrm{2}}$ on the {hBN} {Substrate} {Competing} with {Nonradiative} {Exciton}–{Exciton} {Annihilation}.
\newblock {\em \href{https://pubs.acs.org/doi/10.1021/acsphotonics.1c01584}{ACS Photonics}}, 9(3):873, March 2022.

\bibitem{hoshi2017}
Yusuke Hoshi, Takashi Kuroda, Mitsuhiro Okada, Rai Moriya, Satoru Masubuchi, Kenji Watanabe, Takashi Taniguchi, Ryo Kitaura, and Tomoki Machida.
\newblock Suppression of exciton-exciton annihilation in tungsten disulfide monolayers encapsulated by hexagonal boron nitrides.
\newblock {\em \href{https://link.aps.org/doi/10.1103/PhysRevB.95.241403}{Physical Review B}}, 95(24):241403, June 2017.

\bibitem{Chatterjee2022}
Suman Chatterjee, Garima Gupta, Sarthak Das, Kenji Watanabe, Takashi Taniguchi, and Kausik Majumdar.
\newblock Trion-trion annihilation in monolayer {WS} 2.
\newblock {\em \href{https://link.aps.org/doi/10.1103/PhysRevB.105.L121409}{Physical Review B}}, 105(12):L121409, March 2022.

\bibitem{chaudhary2020}
Raghav Chaudhary, Varun Raghunathan, and Kausik Majumdar.
\newblock Origin of selective enhancement of sharp defect emission lines in monolayer {WSe$_{\textrm{2}}$} on rough metal substrate.
\newblock {\em \href{https://pubs.aip.org/jap/article/127/7/073105/565846/Origin-of-selective-enhancement-of-sharp-defect}{Journal of Applied Physics}}, 127(7):073105, February 2020.

\bibitem{simon_quantum_2010}
C.~Simon, M.~Afzelius, J.~Appel, A.~Boyer De La~Giroday, S.~J. Dewhurst, N.~Gisin, C.~Y. Hu, F.~Jelezko, S.~Kröll, J.~H. Müller, J.~Nunn, E.~S. Polzik, J.~G. Rarity, H.~De~Riedmatten, W.~Rosenfeld, A.~J. Shields, N.~Sköld, R.~M. Stevenson, R.~Thew, I.~A. Walmsley, M.~C. Weber, H.~Weinfurter, J.~Wrachtrup, and R.~J. Young.
\newblock Quantum memories: {A} review based on the {European} integrated project “{Qubit} {Applications} ({QAP})”.
\newblock {\em \href{http://link.springer.com/10.1140/epjd/e2010-00103-y}{The European Physical Journal D}}, 58(1):1, May 2010.

\bibitem{gao_atomically-thin_2023}
Timm Gao, Martin Von~Helversen, Carlos Antón-Solanas, Christian Schneider, and Tobias Heindel.
\newblock Atomically-thin single-photon sources for quantum communication.
\newblock {\em \href{https://www.nature.com/articles/s41699-023-00366-4}{npj 2D Materials and Applications}}, 7(1):4, January 2023.

\bibitem{peng2020}
Lintao Peng, Henry Chan, Priscilla Choo, Teri~W. Odom, Subramanian K. R.~S. Sankaranarayanan, and Xuedan Ma.
\newblock Creation of {Single}-{Photon} {Emitters} in {WSe}$_{\textrm{2}}$ {Monolayers} {Using} {Nanometer}-{Sized} {Gold} {Tips}.
\newblock {\em \href{https://pubs.acs.org/doi/10.1021/acs.nanolett.0c01789}{Nano Letters}}, 20(8):5866, August 2020.

\bibitem{wang2021}
Qixing Wang, Julian Maisch, Fangdong Tang, Dong Zhao, Sheng Yang, Raphael Joos, Simone~Luca Portalupi, Peter Michler, and Jurgen~H. Smet.
\newblock Highly {Polarized} {Single} {Photons} from {Strain}-{Induced} {Quasi}-{1D} {Localized} {Excitons} in {WSe}$_{\textrm{2}}$.
\newblock {\em \href{https://pubs.acs.org/doi/10.1021/acs.nanolett.1c01927}{Nano Letters}}, 21(17):7175, September 2021.

\bibitem{shepard2017}
Gabriella~D Shepard, Obafunso~A Ajayi, Xiangzhi Li, X-Y Zhu, James Hone, and Stefan Strauf.
\newblock Nanobubble induced formation of quantum emitters in monolayer semiconductors.
\newblock {\em \href{https://iopscience.iop.org/article/10.1088/2053-1583/aa629d}{2D Materials}}, 4(2):021019, March 2017.

\bibitem{Srivastava2015}
Ajit Srivastava, Meinrad Sidler, Adrien~V. Allain, Dominik~S. Lembke, Andras Kis, and A.~Imamoğlu.
\newblock Optically active quantum dots in monolayer {WSe$_{\textrm{2}}$}.
\newblock {\em \href{https://www.nature.com/articles/nnano.2015.60}{Nature Nanotechnology}}, 10(6):491, June 2015.

\bibitem{Rosenberger2019}
Matthew~R. Rosenberger, Chandriker~Kavir Dass, Hsun~Jen Chuang, Saujan~V. Sivaram, Kathleen~M. McCreary, Joshua~R. Hendrickson, and Berend~T. Jonker.
\newblock {Quantum Calligraphy: Writing Single-Photon Emitters in a Two-Dimensional Materials Platform}.
\newblock {\em \href{https://pubs.acs.org/doi/10.1021/acsnano.8b08730}{ACS Nano}}, 13(1):904, 2019.

\bibitem{hernandez_lopez_strain_2022}
Pablo Hernández~López, Sebastian Heeg, Christoph Schattauer, Sviatoslav Kovalchuk, Abhijeet Kumar, Douglas~J. Bock, Jan~N. Kirchhof, Bianca Höfer, Kyrylo Greben, Denis Yagodkin, Lukas Linhart, Florian Libisch, and Kirill~I. Bolotin.
\newblock Strain control of hybridization between dark and localized excitons in a {2D} semiconductor.
\newblock {\em \href{https://www.nature.com/articles/s41467-022-35352-9}{Nature Communications}}, 13(1):7691, December 2022.

\bibitem{abramov_photoluminescence_2023}
Artem~N. Abramov, Igor~Y. Chestnov, Ekaterina~S. Alimova, Tatiana Ivanova, Ivan~S. Mukhin, Dmitry~N. Krizhanovskii, Ivan~A. Shelykh, Ivan~V. Iorsh, and Vasily Kravtsov.
\newblock Photoluminescence imaging of single photon emitters within nanoscale strain profiles in monolayer {WSe$_{\textrm{2}}$}.
\newblock {\em \href{https://www.nature.com/articles/s41467-023-41292-9}{Nature Communications}}, 14(1):5737, September 2023.

\bibitem{moon_strain-correlated_2020}
Hyowon Moon, Eric Bersin, Chitraleema Chakraborty, Ang-Yu Lu, Gabriele Grosso, Jing Kong, and Dirk Englund.
\newblock Strain-{Correlated} {Localized} {Exciton} {Energy} in {Atomically} {Thin} {Semiconductors}.
\newblock {\em \href{https://pubs.acs.org/doi/10.1021/acsphotonics.0c00626}{ACS Photonics}}, 7(5):1135, May 2020.

\bibitem{xu_conversion_2023}
David~D. Xu, Albert~F. Vong, Dmitry Lebedev, Riddhi Ananth, Alexa~M. Wong, Paul~T. Brown, Mark~C. Hersam, Chad~A. Mirkin, and Emily~A. Weiss.
\newblock Conversion of {Classical} {Light} {Emission} from a {Nanoparticle}‐{Strained} {WSe}$_{\textrm{2}}$ {Monolayer} into {Quantum} {Light} {Emission} via {Electron} {Beam} {Irradiation}.
\newblock {\em \href{https://onlinelibrary.wiley.com/doi/10.1002/adma.202208066}{Advanced Materials}}, 35(5):2208066, February 2023.

\bibitem{linhart_localized_2019}
Lukas Linhart, Matthias Paur, Valerie Smejkal, Joachim Burgdörfer, Thomas Mueller, and Florian Libisch.
\newblock Localized {Intervalley} {Defect} {Excitons} as {Single}-{Photon} {Emitters} in {WSe}$_{\textrm{2}}$.
\newblock {\em \href{https://link.aps.org/doi/10.1103/PhysRevLett.123.146401}{Physical Review Letters}}, 123(14):146401, September 2019.

\end{thebibliography}
\newpage
\newpage
\newpage
\newpage
\includepdf[pages=-]{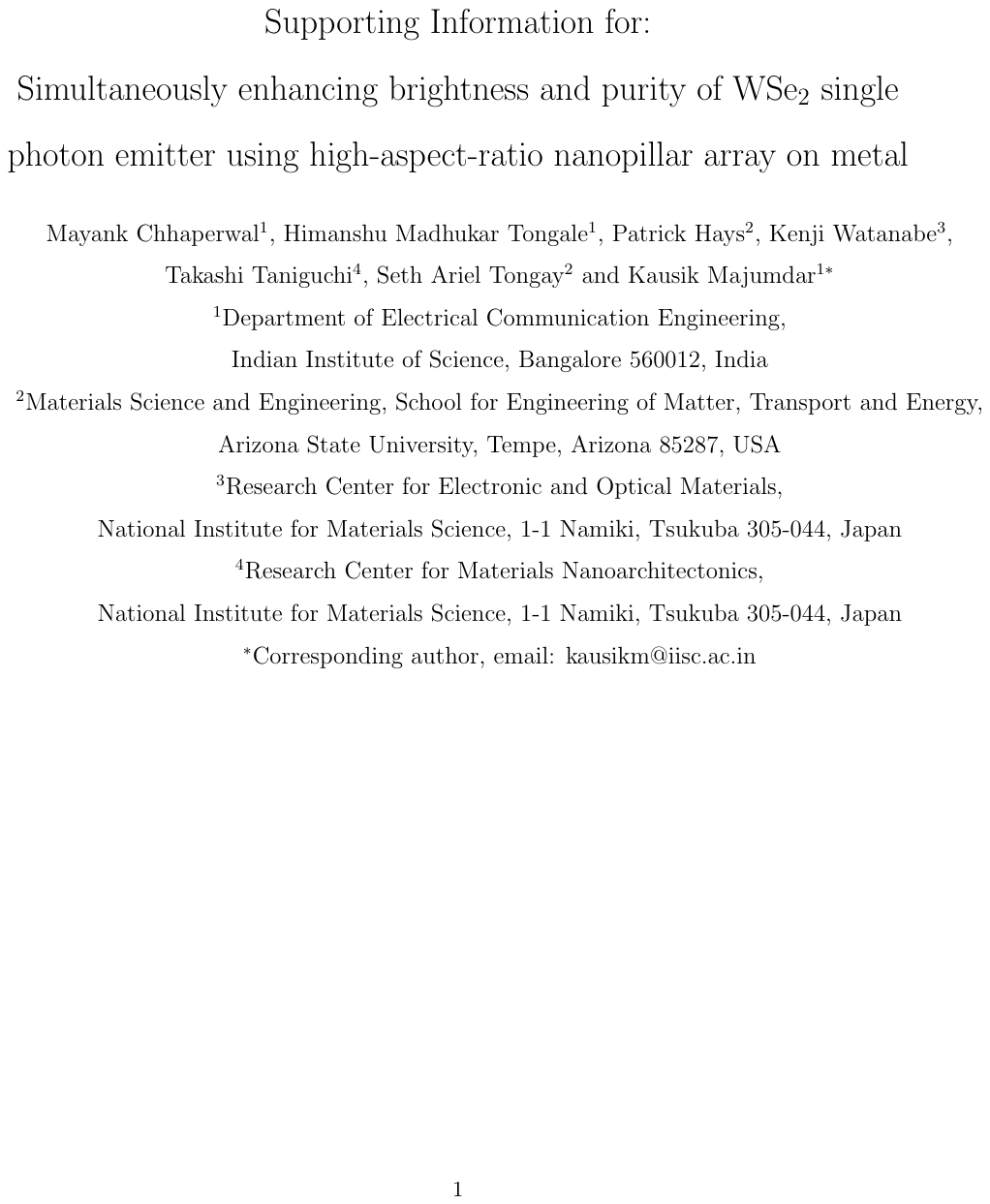}
\end{document}